\documentclass{jnmp}

\def\today{4~December~2008} 

\def \D {\hbox{d}}

\def \sn  {\mathop{\rm sn}\nolimits}
\def \cn  {\mathop{\rm cn}\nolimits}
\def \dn  {\mathop{\rm dn}\nolimits}
\def \sd  {\mathop{\rm sd}\nolimits}
\def \cd  {\mathop{\rm cd}\nolimits}
\def \nd  {\mathop{\rm nd}\nolimits}

\def \ha  {\mathop{\rm h}\nolimits}

\def \mod#1{\vert #1 \vert}
\def \sech{\mathop{\rm sech}\nolimits}

\JNMPnumberwithin{equation}{section}

\theoremstyle{definition}

\begin{document}

%
\renewcommand{\evenhead}{Robert Conte and K.W.~Chow}
\renewcommand{\oddhead}
 {Doubly periodic waves of a discrete NLS system}

%
\thispagestyle{empty}

\FirstPageHead{2008}{*}{*}{\pageref{firstpage}--\pageref{lastpage}}{Article}

\copyrightnote{2008}{Robert Conte and K.W.~Chow}

\Name
 {Doubly periodic waves of a discrete nonlinear Schr\"odinger system
  with saturable nonlinearity}

\label{firstpage}

\Author{Robert CONTE}

\Address{
LRC MESO,
\\ \'Ecole normale sup\'erieure de Cachan (CMLA) et CEA--DAM
\\ 61, avenue du Pr\'esident Wilson,
\\ F--94235 Cachan Cedex, France.
\\and
\\ Department of Mathematics, University of Hong Kong
\\ Pokfulam, Hong Kong
\\ E-mail:  Robert.Conte@cea.fr
}

\Author{K.W.~Chow}

\Address{
Department of Mechanical Engineering, University of Hong Kong
\\ Pokfulam, Hong Kong
\\ E-mail: kwchow@hkusua.hku.hk
}

\Date{Received 30~September~2008; Revised \today; Accepted 4 December 2008}

\begin{abstract}
A system of two discrete
nonlinear Schr\"odinger equations
of the Ablowitz-Ladik type
with a saturable nonlinearity
is shown to admit a doubly periodic wave,
whose long wave limit is also derived.
As a by-product,
several new solutions of the elliptic type are provided
for NLS-type discrete and continuous systems.
\end{abstract}



\noindent \textit{PACS 2001} 
  02.30.Jr,
  02.30.Ik,
  42.65.Wi.


\section{Introduction}
\label{sectionIntroduction}

The discrete nonlinear Schr\"odinger equation
(d-NLS)
\begin{eqnarray}
& &
{\hskip -5.0 truemm}
 i A_t +  p \frac{A(x+h,t)+A(x-h,t)-2 A(x,t)}{h^2}
 + q {\mod A}^2 A=0,
p q \not=0,\
p \hbox{ and } q \hbox{ real},
\label{eqdNLS_nonAL}
\end{eqnarray}
occurs in many physical disciplines, but unfortunately is not
integrable. 
Hence most investigations must be conducted via
numerical computations or other means of approximation.
In contrast, the Ablowitz--Ladik (AL) model \cite{AblowitzLadik},
\begin{eqnarray}
& &
{\hskip -18.0 truemm}
 i A_t +  p \frac{A(x+h,t)+A(x-h,t)-2 A(x,t)}{h^2}
 + q \frac{A(x+h,t)+A(x-h,t)}{2} \mod{A(x,t)}^2=0,
\label{eqdNLS_AL}
\end{eqnarray}
possesses $N$-soliton solutions, periodic waves, 
an infinite number of conservation laws 
and a host of other properties associated with
integrable systems \cite{V1999,Z2000}. 
Although d-NLS occurs much more frequently than AL in physical situations, 
there are still some situations where AL is relevant, 
e.g., 
higher order evolution equations \cite{H1998}, 
optics \cite{MFK2006}, 
and protein dynamics \cite{K2000}. 
Indeed generalizations of AL systems have received quite intensive 
attention in the literature \cite{P2006}.

The goal here is to investigate a class of
coupled (or vector) Ablowitz--Ladik systems with saturable
nonlinearities. 
By ``saturable'' NLS-type equation,
we mean that, in the nonlinear term,
the coefficient of $A$, typically $\mod{A}^2$,
is replaced by some expression whose limit is finite
when $\mod{A} \to + \infty$, such as
$\mod{A}^2/(1+\mod{B}^2)$
or, in the case of two-component NLS,
by 
$\mod{A}^2/(\mod{A}^2+\mod{A}^2)$.
A brief and certainly incomplete review
of some existing works in the literature will provide the
motivation of the present work.

We first look at recent advances on the AL system.
Straightforward AL extensions of vector nonlinear Schr\"{o}dinger
equations exhibit solitons with elastic collision properties \cite{AOT}.
Indeed collision properties of chains of near identital solitons
have been studied too \cite{D2003,D2004}.
Other variants, e.g. coupled discrete evolution systems with
nonlinearities resembling the AL model \cite{B1999,MY2002},
have been considered. 
However, they are usually not integrable, 
and propagation of solitons may be unstable.

A remark on saturable nonlinearity is also in order.
Saturable nonlinearities probably first arose in the setting
of nonlinear optics. 
In an optical fiber, the Kerr nonlinearity typically generates the
conventional nonlinear Schr\"odinger equation (NLS).
However,
for short pulses and a high input peak pulse power,
these Kerr nonlinearities cannot adequately describe
the field-induced change in the refractive index.
In such a case, it is sufficient to introduce a saturation term
in the nonlinearity, leading in the simplest case to
the one-component saturable (continuous) nonlinear Schr\"odinger equation
\cite{GH1991}
\begin{eqnarray}
& &
{\hskip -5.0 truemm}
 i A_t +  p A_{xx} + q \frac{ {\mod A}^2 A}{1 + \mu {\mod A}^2}=0,
\label{eqcNLS_saturable}
\end{eqnarray}
in which $p, q, \mu$ are real constants.
Other types of nonlinearities have also been considered
\cite{KLMB}.

Its d-NLS-type discretization
\begin{eqnarray}
& &
{\hskip -5.0 truemm}
 i A_t +  p \frac{A(x+h,t)+A(x-h,t)-2 A(x,t)}{h^2}
 + q \frac{ {\mod A}^2 A}{1 + \nu (q h^2/p) {\mod A}^2}=0,
p q \nu \not=0,
\label{eqdNLS_nonAL_saturable}
\end{eqnarray}
in which the saturation coefficient $\nu$ is dimensionless,
has a considerable importance in nonlinear optics
\cite{Efremidis_etalPRE2002},
but very few analytic results are known.
Apart from plane waves and some perturbative solution \cite{SKHM2004},
the main particular solution so far
is an elliptic stationary wave solution \cite{KRSS},
recalled as Eq.~(\ref{eqdNLS_nonAL_saturable_elliptic_Khare})
in section \ref{sectionOne_saturable_discrete_NLS}.
In particular the problem remains open to find an exact travelling wave of 
(\ref{eqdNLS_nonAL_saturable})
with an arbitrary velocity.

When the saturation term $1/(1+ \nu (q h^2/p) {\mod A}^2)$
is replaced by an arbitrary analytic function of ${\mod A}^2$
and, at the same time, the nonlinearity is partly discretized
in the manner of Ablowitz and Ladik,
\begin{eqnarray}
& &
 i A_t +  p \frac{A(x+h,t)+A(x-h,t)-2 A(x,t)}{h^2}
\nonumber \\ & & \phantom{123}
 + q \left(F(\mod A) \frac{A(x+h,t)+A(x-h,t)}{2}
          +G(\mod A) A(x,t)\right)=0,
\label{eqdNLS_mixedAL}
\end{eqnarray}
one can require the existence of 
a special class of moving kinks and pulses 
and thus determine $F$ and $G$ \cite{FZK1999},
with the main result that Eq.~(\ref{eqdNLS_nonAL_saturable})
admits no stationary pulse in the considered class.
A similar inverse approach has been followed in Ref.~\cite{Kh2006}.

Over the years, dynamics of pulses in saturable media
has continued to attract attention. 
An example of a coupled waveguide,
which we plan to expand on here, focuses on a saturable,
nonlinear medium with cross phase modulation. 
The propagation of
two-component spatial optical solitons carrying angular momentum
has been investigated \cite{SK2004}.
The reduction of this system to one transverse dimension
defines the two-component saturable NLS
\begin{eqnarray}
& &
\left\lbrace
\begin{array}{ll}
\displaystyle{
 i A_t +  \lambda_1 p A_{xx}
 + q \frac{ {\mod A}^2 + \sigma_1{\mod B}^2}
          {1 + s_1 \left({\mod A}^2+{\mod B}^2\right)} A=0,
} \\ \displaystyle{
 i B_t +  \lambda_2 p B_{xx}
 + q \frac{ {\mod B}^2 + \sigma_2 {\mod A}^2}
          {1 + s_2 \left({\mod A}^2+{\mod B}^2\right)} B=0,
}
\end{array}
\right.
\label{eqcNLS_saturable_two}
\end{eqnarray}
in which $p, q, \lambda_1, \lambda_2, s_1, s_2, \sigma_1, \sigma_2$ 
are real constants,
and this system
has been shown \cite{CMN2006} to admit at least one stationary solution
(of elliptic type)
with a nonconstant total intensity $\mod{A}^2 +\mod{B}^2$,
so that the system is truly saturable.
In section \ref{sectionTwo_saturable_discrete_NLS_AL},
we extrapolate this solution to a travelling wave having an arbitrary velocity.
Another two-component saturable NLS,
\begin{eqnarray}
& &
\left\lbrace
\begin{array}{ll}
\displaystyle{
 i A_t +  \lambda_1 p A_{xx}
 + q \frac{A}
          {1 + s_1 \left({\mod A}^2+{\mod B}^2\right)}=0,
} \\ \displaystyle{
 i B_t +  \lambda_2 p B_{xx}
 + q \frac{B}
          {1 + s_2 \left({\mod A}^2+{\mod B}^2\right)}=0,
}
\end{array}
\right.
\label{eqcNLS_saturable_two_variant}
\end{eqnarray}
describes optically induced nonlinear photonic lattices
in which discrete solitons have been observed
\cite{Efremidis_etalPRE2002,Fleischer_etalNature2003}.

{}From the perspective of nonlinear science, 
as well as from the viewpoint of applications, 
it is therefore instructive to consider NLS-type equations
which are discrete in space and possess a saturable nonlinearity.
Given the elegant nature of the AL system, it seems plausible
to study such a discrete saturable NLS
with the nonlinear term discretized in the manner of AL rather than
of the d-NLS.
In fact preliminary studies of such systems have already
begun \cite{KRSS,Kh2006}.

The structure of this paper is the following. 
In section \ref{sectionSimplificationJacobi}
we explain how to perform computations which minimize the algebraic
complexity.
In section \ref{sectionOne_saturable_discrete_NLS}, 
we first
describe a new family of stationary solutions for a single component AL
with saturable nonlinearity.
These new solutions are different from those
given in the literature \cite{KRSS,Kh2006}. 
They are delineated by a cyclic
combination of elliptic functions in the compact and
efficient notation due to Halphen. 

Then in section \ref{sectionTwo_saturable_continuous_NLS},
we consider a system of two coupled saturable continuous NLS
and generalize a structure previously obtained \cite{CMN2006}
to an arbitrary velocity.

Finally in section 
\ref{sectionTwo_saturable_discrete_NLS_AL}
we introduce an AL-type
discretization of system (\ref{eqcNLS_saturable_two})
and find exact solutions with a nonconstant total intensity
${\mod A}^2+{\mod B}^2$,
so that the saturability feature is effective.
These solutions are stationary and of the elliptic type.
To help the reader,
we have gathered in Appendix all the mentioned solutions, 
whether existing or new.

\section{Practical computation}
\label{sectionSimplificationJacobi}

Since the twelve Jacobi elliptic functions
\cite[Chap.~16]{AbramowitzStegun}
are equivalent in the complex plane,
any elliptic solution
can be presented as twelve equivalent complex expressions.
However, in physics we are only interested in those expressions which are 
bounded on the real axis,
i.e.~which only involve 
the six functions $\sn,\cn,\dn$ and $\sd,\cd,\nd$
of the canonical argument $(Q x,k)$,
in which the Jacobi modulus $k$ lies between $0$ and $1$
and the Jacobi nome $Q$ is positive 
\cite[\S 18.9.9, 18.9.12]{AbramowitzStegun},
\begin{eqnarray}
& &
m=k^2=\frac{e_2-e_3}{e_1-e_3},\ Q^2=e_1-e_3,\
0<m<1,\ 0<Q^2,\ 
e_3 < e_2 < e_1. 
\label{eqAS18.9.9}
\end{eqnarray}

In order to minimize time and effort,
we will follow a two-step procedure.
\begin{enumerate}
\item
Firstly, to find a complex (mathematical) solution 
$A=f(\ha_\alpha(x),\ha_\beta(x),\ha_\gamma(x))$,
in which 
$(\alpha,\beta,\gamma)$ is an arbitrary permutation of $(1,2,3)$
and $(\ha_1,\ha_2,\ha_3)$ 
are the elliptic functions introduced by Halphen \cite{HalphenTraite},
defined by the differential system
\begin{eqnarray}
& &
\frac{\D \ha_\alpha}{\D x}= - \ha_\beta \ha_\gamma,\
\ha_\beta^2-\ha_\gamma^2=-e_\beta+e_\gamma,\
e_1+e_2+e_3=0,
\label{eqHalphen_system}
\end{eqnarray}
with the choice of sign $\lim_{x \to 0} x \ha_\alpha(x)=+1$.
The invariance under any permutation of $(\alpha,\beta,\gamma)$
is an important practical advantage of the Halphen notation
over the Jacobi notation.

\item
Secondly, for each such complex solution,
to convert the Halphen trio $(\ha_1,\ha_2,\ha_3)$
to both bounded Jacobi trios 
$(\sn,\cn,\dn)$ and $(\sd,\cd,\nd)$
by formulae such as
\begin{eqnarray}
& &
\left\lbrace
\begin{array}{ll}
\displaystyle{
\ha_1(x-x_0) = a_1 \dn(Q x,k),\
\ha_2(x-x_0) = a_2 \cn(Q x,k),\
} \\ \displaystyle{
\ha_3(x-x_0) = a_3 \sn(Q x,k),\
} \\ \displaystyle{
a_1^2=e_3-e_1,\
a_2^2=e_3-e_2,\
a_3^2=e_2-e_3,\
\frac{a_2 a_3}{a_1}=Q k^2,
}
\end{array}
\right.
\label{eqHalphenJacobi_sncndn}
\end{eqnarray}
in which $x_0$ is an immaterial complex shift.
Full details can be found in Ref.~\cite[App.~B]{CCX}.

\end{enumerate}

\section{One-component saturable discrete NLS}
\label{sectionOne_saturable_discrete_NLS}

For convenience, let us introduce the constant $a_0$,
a kind of unit of amplitude,
\begin{eqnarray}
& &
a_0^2=-2 \frac{p}{q}.
\end{eqnarray}

Let us first consider Eq.~(\ref{eqdNLS_nonAL_saturable}),
\begin{eqnarray}
& &
{\hskip -5.0 truemm}
 i A_t +  p \frac{A(x+h,t)+A(x-h,t)-2 A(x,t)}{h^2}
 + q \frac{ {\mod A}^2 A}{1 + \nu (q h^2/p) {\mod A}^2}=0,
p q \nu \not=0.
{\hskip 0.2 truemm} (\ref{eqdNLS_nonAL_saturable})
\nonumber
\end{eqnarray}

If one assumes for a solution of (\ref{eqdNLS_nonAL_saturable})
the expression
(first step as defined in section \ref{sectionSimplificationJacobi})
\begin{eqnarray}
& &
A=a_1 \left\lbrack
 \ha_\alpha(x-ct) + i b_1 \frac{\ha_\beta(x-ct)}{\ha_\gamma(x-ct)}
      \right\rbrack
e^{i(K (x-ct) - \omega t)},\
\label{eqdNLS_nonAL_saturable_elliptic_Halphen}
\end{eqnarray}
one finds six constraints among
the seven introduced constants $(a_1,b_1,c,\omega,K,e_\alpha,e_\beta)$,
\begin{eqnarray}
& & \left\lbrace
\begin{array}{ll}
\displaystyle{
c=0,\
\omega=\frac{p}{h^2} \left(2 - \frac{1}{\nu} \right),\
\tan K h =- b_1 \frac{\ha_\beta(h)}{\ha_\alpha(h) \ha_\gamma(h)},
} \\ \displaystyle{
b_1 (e_\gamma-e_\beta) (e_\gamma-e_\alpha - b_1^2)=0,\
} \\ \displaystyle{
} \\ \displaystyle{
a_1^2=a_0^2 \frac{\ha_\gamma^3(h)}
 {4 \nu^2 h^2 \ha_\beta(h) \cos K h
     (b_1^2 (e_\beta-e_\gamma) + \ha_\gamma^4(h))},\
} \\ \displaystyle{
} \\ \displaystyle{
2 \nu = \frac
 {\ha_\gamma(h)
  \left[b_1^2(\ha_\alpha^2(h) \ha_\gamma^4(h)
              -(e_\gamma-e_\alpha)\ha_\beta^4(h))
        +\ha_\alpha^2(h) \ha_\gamma^6(h) \right]}
 {\ha_\beta(h) \cos K h
  \left(b_1^2 (e_\beta-e_\gamma) + \ha_\gamma^4(h)\right)^2}.
}
\end{array}
\right.
\label{eqdNLS_nonAL_saturable_elliptic_Halphen_3sol}
\end{eqnarray}
The factorized form of the fourth constraint
(quite similar to that obtained in \cite{ConteChow2006})
defines three solutions,
each depending on one arbitrary constant
and requiring that $\nu$ be nonzero.
In order to write them explicitly, let us introduce the nonzero constant
\begin{eqnarray}
& &
N_\alpha=\ha_\alpha^4(h) - (e_\alpha-e_\beta)(e_\alpha-e_\gamma)
= 2 \ha_\alpha(h) \ha_\beta(h) \ha_\gamma(h) \ha_\alpha(2 h).
\end{eqnarray}
These three solutions are
\begin{description}
\item $\circ$
$(b_1=0)$:
the elliptic solution already obtained \cite{KRSS},
\begin{eqnarray}
& & 
A=a_1 \ha_\alpha(x) e^{- i \omega t},\
a_1^2=a_0^2 \frac{\ha_\beta(h) \ha_\gamma(h)}{h^2 \ha_\alpha^4(h)},\
2 \nu = \frac{\ha_\alpha^2(h)} {\ha_\beta(h) \ha_\gamma(h)},\
\label{eqdNLS_nonAL_saturable_elliptic_Khare}
\end{eqnarray}

\item $\circ$
$(e_\gamma-e_\beta=0)$:
a dark one-soliton solution, 
\begin{eqnarray}
& & {\hskip -10.0 truemm}
\left\lbrace
\begin{array}{ll}
\displaystyle{
A=a_1 
  \left\lbrack \frac{\kappa}{2} \tanh \frac{\kappa}{2} x + i b_1 \right\rbrack
  e^{i(K x - \omega t)},\
\tan K h= - b_1 \frac{2}{\kappa} \tanh \frac{\kappa h}{2} 
} \\ \displaystyle{
a_1^2=a_0^2 \left(\frac{2}{\kappa h}\right)^2 \sinh^2 \frac{\kappa h}{2}
            \cosh^{-4} \frac{\kappa h}{2} \cos^5 K h,\
2 \nu = \cosh^{2} \frac{\kappa h}{2} \cos^{-3} K h,\
}
\end{array}
\right.
\label{eqdNLS_nonAL_saturable_dark}
\end{eqnarray}

\item $\circ$
$(e_\gamma-e_\alpha-b_1^2=0)$:
a new elliptic solution,
different from the previous one,
\begin{eqnarray}
& &
\left\lbrace
\begin{array}{ll}
\displaystyle{
A=a_1 \left\lbrack
 \ha_\alpha(x) + i b_1 \frac{\ha_\beta(x)}{\ha_\gamma(x)}
      \right\rbrack
e^{i(K x - \omega t)},\
b_1^2=e_\gamma-e_\alpha,\
} \\ \displaystyle{
} \\ \displaystyle{
a_1^2=a_0^2 \frac{\ha_\beta(h) \ha_\gamma(h) N_\gamma}
                 {h^2 N_\alpha^2 \cos K h},\
2 \nu=\frac{\ha_\gamma(h) N_\alpha}
         {\ha_\beta(h) N_\gamma \cos K h}.
}
\end{array}
\right.
\label{eqdNLS_nonAL_saturable_elliptic_Halphen_sole}
\end{eqnarray}

\end{description}

In the continuum limit,
the coefficient $\nu$ goes to $1/2$ for all three solutions,
so 
the limit for $\omega$ requires expanding $\nu$ to second order in $h$.
Each solution goes to some solution of the continuous NLS,
\begin{description}
\item $\circ$
$(b_1=0)$:
\begin{eqnarray}
& & 
e_\beta,e_\gamma=\hbox{arbitrary},\
A=a_0 \ha_\alpha(x) e^{- i \omega t},\
2 \nu = 1,\
\omega=- 3 p e_\alpha,\
\label{eqcNLS_elliptic_Lame}
\end{eqnarray}

\item $\circ$
$(e_\gamma-e_\beta=0)$:
\begin{eqnarray}
& & \left\lbrace
\begin{array}{ll}
\displaystyle{
b_1,\kappa=\hbox{arbitrary},\
A=a_0 \left\lbrack \frac{\kappa}{2} \tanh \frac{\kappa}{2} x + i b_1 
     \right\rbrack
 e^{i(-b_1 x - \omega t)},\
} \\ \displaystyle{
2 \nu = 1,\
\omega=\left(3 b_1^2+\frac{\kappa^2}{2}\right) p,\
}
\end{array}
\right.
\label{eqcNLS_dark}
\end{eqnarray}

\item $\circ$
$(e_\gamma-e_\alpha-b_1^2=0)$:

\begin{eqnarray}
& & {\hskip -20.0 truemm} \left\lbrace
\begin{array}{ll}
\displaystyle{
e_\beta,e_\gamma=\hbox{arbitrary},\
A=a_0 \left\lbrack
 \ha_\alpha(x) + i b_1 \frac{\ha_\beta(x)}{\ha_\gamma(x)}
      \right\rbrack
e^{i(-b_1 x - \omega t)},\
b_1^2=e_\gamma-e_\alpha,\
} \\ \displaystyle{
2 \nu = 1,\
\omega=(4 e_\beta + 5 e_\gamma) p.
}
\end{array}
\right.
\label{eqcNLS_elliptic-order-4}
\end{eqnarray}

\end{description}

Following the second step of section \ref{sectionSimplificationJacobi},
various bounded solutions represented by the complex expression
(\ref{eqdNLS_nonAL_saturable_elliptic_Halphen_sole})
can be found in terms of bounded Jacobi functions,
such as
\begin{eqnarray}
& &
A=
(\sn+i \cd) e^{i(K x -\omega t)},\
(\cn+i \sd) e^{i(K x -\omega t)},\
\dots
\end{eqnarray}
we leave this exercise to the interested reader.

On this NLS limit,
it is easier to give the precise difference between the two elliptic solutions
(\ref{eqcNLS_elliptic_Lame}) and 
(\ref{eqcNLS_elliptic-order-4}),
i.e.~between
(\ref{eqdNLS_nonAL_saturable_elliptic_Khare}) and 
(\ref{eqdNLS_nonAL_saturable_elliptic_Halphen_sole}).
Indeed, the travelling wave reduction of NLS
\begin{eqnarray}
& &
A(x,t)=\sqrt{M(\xi)} e^{i(\displaystyle{-\omega t + \varphi(\xi)})},\
\xi=x-ct,
\label{eqCGL3red}
\end{eqnarray}
admits as general solution the elliptic functions
\begin{eqnarray}
& &
\left\lbrace
\begin{array}{ll}
\displaystyle{
M=\mod{A}^2=a_0^2 \left(\wp(\xi-\xi_0,g_2,g_3) - e_0\right),\
\varphi' = \frac{c}{2 p} + \frac{K_1}{\mod{A}^2},\
a_0^2=-2 \frac{p}{q},\
} \\ \displaystyle{
e_0=\frac{4 \omega p - c^2}{12 p^2},\
K_1^2=- \frac{a_0^4}{4} (4 e_0^3-g_2 e_0-g_3), 
}
\end{array}
\right.
\label{eqNLSmod2argprime}
\end{eqnarray}
therefore the above two elliptic solutions can only be particular cases of it.
The first solution
(\ref{eqcNLS_elliptic_Lame}) 
is the well known Lam\'e function corresponding to $e_0=e_\alpha,K_1=0$.
As to the second solution
(\ref{eqcNLS_elliptic-order-4}),
by establishing the ODEs for $\mod{A}^2$ and $(\arg A)'$
one finds the values 
\begin{eqnarray}
& &
e_0=3 E_\alpha+E_\beta,\
K_1= -2 a_0^2 \sqrt{E_\gamma-E_\alpha} (E_\alpha-E_\beta),
\end{eqnarray}
in which the elliptic invariants $E_j$ of (\ref{eqcNLS_elliptic-order-4})
are related to those $e_j$ of (\ref{eqcNLS_elliptic_Lame}) 
by the Landen transformation which divides by two only one of the two periods
\cite[\S 13.23]{Bateman},
\begin{eqnarray}
& &
\wp(z|\omega,\omega') \to \wp\Big(z\Big|\omega,\frac{\omega'}{2}\Big).
\end{eqnarray}
This is the nonzero value of $K_1$ which proves that both solutions
are indeed different.

We have checked that the saturable discrete NLS (\ref{eqdNLS_nonAL_saturable})
does not admit a solution depending on two elliptic functions
with different periods such as the one \cite{CCX}
admitted by the discrete integrable NLS of Ablowitz and Ladik.

\section{Two-component saturable continuous NLS}
     \label{sectionTwo_saturable_continuous_NLS}

We next consider the two-component saturable continuous NLS 
Eq.~(\ref{eqcNLS_saturable_two}),
\begin{eqnarray}
& &
\left\lbrace
\begin{array}{ll}
\displaystyle{
 i A_t +  \lambda_1 p A_{xx}
 + q \frac{ {\mod A}^2 + \sigma_1{\mod B}^2}
          {1 + s_1 \left({\mod A}^2+{\mod B}^2\right)} A=0,
} \\ \displaystyle{
 i B_t +  \lambda_2 p B_{xx}
 + q \frac{ {\mod B}^2 + \sigma_2 {\mod A}^2}
          {1 + s_2 \left({\mod A}^2+{\mod B}^2\right)} B=0,
}
\end{array}
\right.
{\hskip 42.0 truemm} (\ref{eqcNLS_saturable_two})
\nonumber
\end{eqnarray}
In the continuum case,
when one requires the total intensity $\mod{A}^2 +\mod{B}^2$ to be nonconstant,
so that the system (\ref{eqcNLS_saturable_two}) truly display saturability,
then only one elliptic solution is known \cite{CMN2006},
in which the moduli $\mod{A}, \mod{B}$ have a zero velocity $c$.
However, if one assumes
\begin{eqnarray}
& &
\left\lbrace
\begin{array}{ll}
\displaystyle{
A=a_1 \ha_\beta(x-ct) \ha_\gamma(x-ct)
      e^{\displaystyle{i(K_a (x-ct)-\omega_a t)}},
} \\ \displaystyle{
B=a_2 \ha_\beta(x-ct) \ha_\alpha(x-ct)
      e^{\displaystyle{i(K_b (x-ct)-\omega_b t)}},\
      e_\gamma \not= e_\alpha,
}
\end{array}
\right.
\label{eqcNLS_saturable_two_product}
\end{eqnarray}
the above mentioned solution,
which contained no arbitrary parameter at all,
is extended to
a new solution with one arbitrary parameter (the velocity $c$),
defined by the algebraic relations
\begin{eqnarray}
& &
\left\lbrace
\begin{array}{ll}
\displaystyle{
c=\hbox{ arbitrary},\
a_1^2= a,\
a_2^2=-a,\
} \\ \displaystyle{
a=9 a_0^4 
\frac{\displaystyle{\frac{\lambda_1}{1-\sigma_1}+\frac{\lambda_2}{1-\sigma_2}}}
     {\displaystyle{\frac{1}{\lambda_1 s_1}-\frac{1}{\lambda_2 s_2}
                -\frac{1}{2}
 \left(\frac{1-\sigma_1}{\lambda_1 s_1}-\frac{1-\sigma_2}{\lambda_2 s_2}\right)}},
} \\ \displaystyle{
e_\gamma-e_\alpha
  =-\frac{1-\sigma_1}{3 a_0^2 \lambda_1 s_1}
  = \frac{1-\sigma_2}{3 a_0^2 \lambda_2 s_2}\not=0,\
} \\ \displaystyle{
e_\gamma+e_\alpha
 = \frac{1-\sigma_1^2}{9 a_0^2 s_1}-2 \frac{a_0^2 \lambda_1^2}{a}
 = \frac{1-\sigma_2^2}{9 a_0^2 s_2}+2 \frac{a_0^2 \lambda_2^2}{a},\
} \\ \displaystyle{
K_a=\frac{c}{2 p \lambda_1},\
K_b=\frac{c}{2 p \lambda_2},\
} \\ \displaystyle{
\omega_a=-\frac{c}{2} K_a + \frac{2 p}{s_1 a_0^2} 
 - 3 p \lambda_1 e_\alpha
 - \frac{18 p \lambda_1^2 a_0^2}{a (1-\sigma_1)},\
} \\ \displaystyle{
\omega_b=-\frac{c}{2} K_b + \frac{2 p}{s_2 a_0^2} 
 - 3 p \lambda_2 e_\gamma
 + \frac{18 p \lambda_2^2 a_0^2}{a (1-\sigma_2)}.
}
\end{array}
\right.
\label{eqcNLS_saturable_two_sol}
\end{eqnarray}
and there exists exactly one constraint among the fixed coefficients,
\begin{eqnarray}
& &
\frac{1-\sigma_1}{\lambda_1 s_1} +\frac{1-\sigma_2}{\lambda_2 s_2}=0,
\label{eqfixed1}
\end{eqnarray}
which implies $s_1 \not= s_2$.
In particular, the dispersion relations take the form
\begin{eqnarray}
& &
\omega_j=-\frac{c}{2} K_j + \hbox{constant}_j,\ j=a,b.
\end{eqnarray}
Since the three roots $e_\alpha$ are always real,
this complex solution
defines three bounded solutions (in terms of $\sn,\cn,\dn$),
identical to those mentioned in \cite[Eqs.~(12)--(13)]{CMN2006},
i.e.~$(A,B)=(\sn \dn, \cn \dn)$,
$(\cn \sn, \dn \sn)$, $(\dn \cn, \sn \cn)$,
with minor adjustments for the additional dependence on $c$.

The long wave limit of this solution is 
$(e_\beta-e_\alpha)(e_\beta-e_\gamma)=0$, i.e.
\begin{eqnarray}
& &
\left\lbrace
\begin{array}{ll}
\displaystyle{
A= \alpha_1 \sech k (x-ct) \tanh k(x-ct)
      e^{\displaystyle{i(K_a (x-ct)-\omega_a t)}},
} \\ \displaystyle{
B= \alpha_2 \sech^2 k(x-ct)
      e^{\displaystyle{i(K_b (x-ct)-\omega_b t)}},
}
\end{array}
\right.
\label{eqcNLS_saturable_two_product_trigo}
\end{eqnarray}
with the relations
\begin{eqnarray}
& &
\left\lbrace
\begin{array}{ll}
\displaystyle{
c=\hbox{ arbitrary},\
\alpha_1^2=\alpha_2^2=\frac{\lambda_1-\lambda_2}{\lambda_2 s_2},\
k^2=\frac{1}{3 a_0^2 s_2} 
 \left(\frac{1}{\lambda_1}-\frac{1}{\lambda_2} \right),
} \\ \displaystyle{
K_a=\frac{c}{2 p \lambda_1},\
K_b=\frac{c}{2 p \lambda_2},\
} \\ \displaystyle{
\omega_a=-\frac{c^2}{4 p \lambda_1} 
 -\frac{2 p}{6 s_2 a_0^2} \left(1-\frac{\lambda_1}{\lambda_2}\right),\ 
\omega_b=-\frac{c^2}{4 p \lambda_2} 
 +\frac{4 p}{3 s_2 a_0^2} \left(1-\frac{\lambda_2}{\lambda_1}\right), 
}
\end{array}
\right.
\label{eqcNLS_saturable_two_sol_trigo}
\end{eqnarray}
and the two fixed constraints 
\begin{eqnarray}
& &
1-\sigma_1=\frac{s_1}{s_2}\left(1-\frac{\lambda_1}{\lambda_2}\right),\
1-\sigma_2=1-\frac{\lambda_2}{\lambda_1}.
\end{eqnarray}

\section{Two-component saturable discrete NLS of A-L type}
\label{sectionTwo_saturable_discrete_NLS_AL}

Let us finally consider an AL-type
discretization of system (\ref{eqcNLS_saturable_two}), namely
\begin{eqnarray}
& &
\left\lbrace
\begin{array}{ll}
\displaystyle{
 i A_t +  \lambda_1 p \frac{A(x+h,t)+A(x-h,t)-2 A(x)}{h^2}
} \\ \displaystyle{\phantom{1234}
 + q \frac{ {\mod A}^2 + \sigma_1{\mod B}^2}
          {1 + s_1 \left({\mod A}^2+{\mod B}^2\right)}
  \frac{A(x+h,t)+A(x-h,t)}{2}=0,
} \\ \displaystyle{
 i B_t +  \lambda_2 p \frac{B(x+h,t)+B(x-h,t)-2 B(x)}{h^2}
} \\ \displaystyle{\phantom{1234}
 + q \frac{ {\mod B}^2 + \sigma_2 {\mod A}^2}
          {1 + s_2 \left({\mod A}^2+{\mod B}^2\right)}
  \frac{B(x+h,t)+B(x-h,t)}{2}=0,
}
\end{array}
\right.
\label{eqdNLS_AL_saturable_two}
\end{eqnarray}
The expression (\ref{eqcNLS_saturable_two_product})
where $x$ is now a discrete variable $x=n h$
is still a solution of (\ref{eqdNLS_AL_saturable_two}),
but with much less freedom among the parameters.
It is defined by
\begin{eqnarray}
& &
{\hskip -16.0 truemm}
\left\lbrace
\begin{array}{ll}
\displaystyle{
c=0,\
a_1^2= a,\
a_2^2=-a,\
K_a=0,\ K_b=0,\
} \\ \displaystyle{
s_1 = +\frac{1}{a (e_\alpha-e_\gamma) (\ha_\gamma^2(h) -3 e_\alpha)},\
\lambda_1 s_1=\frac
     {h^2 \ha_\beta^4(h)}
     {a_0^2 \ha_\alpha^2(h) (3 \ha_\alpha^2(h) + e_\alpha-e_\beta)},\
} \\ \displaystyle{
s_2 = -\frac{1}{a (e_\gamma-e_\alpha) (\ha_\alpha^2(h) -3 e_\gamma)},\
\lambda_2 s_2=\frac
     {h^2 \ha_\beta^4(h)}
     {a_0^2 \ha_\gamma^2(h) (3 \ha_\gamma^2(h) + e_\gamma-e_\beta)},\
} \\ \displaystyle{
} \\ \displaystyle{
} \\ \displaystyle{
\omega_a=-\frac
 {2 p a (e_\gamma-e_\alpha) \ha_\beta(h)(\ha_\gamma^2(h) -3 e_\alpha) 
(\ha_\beta^2(h)(\ha_\beta(h)-\ha_\gamma(h))+2(e_\alpha-e_\beta)\ha_\gamma(h))}
 {a_0^2 \ha_\alpha^2(h) (3 \ha_\alpha^2(h) + e_\alpha-e_\beta)},\
} \\ \displaystyle{
\omega_b=+\frac
 {2 p a (e_\alpha-e_\gamma) \ha_\beta(h)(\ha_\alpha^2(h) -3 e_\gamma) 
(\ha_\beta^2(h)(\ha_\beta(h)-\ha_\alpha(h))+2(e_\gamma-e_\beta)\ha_\alpha(h))}
 {a_0^2 \ha_\gamma^2(h) (3 \ha_\gamma^2(h) + e_\gamma-e_\beta)},\
} \\ \displaystyle{
1-\sigma_1=\frac{(e_\gamma-e_\alpha) (\ha_\alpha^2(h) -e_\alpha+e_\gamma)}
                  {\ha_\alpha^2(h) (3 \ha_\alpha^2(h) +e_\alpha-e_\beta)},\
1-\sigma_2=\frac{(e_\alpha-e_\gamma) (\ha_\gamma^2(h) -e_\gamma+e_\alpha)}
                  {\ha_\gamma^2(h) (3 \ha_\gamma^2(h) +e_\gamma-e_\beta)}.
}
\end{array}
\right.
\label{eqdNLS_AL_saturable_two_sol}
\end{eqnarray}
The velocity $c$ now vanishes
(while it is arbitrary in the continuous case)
and the fixed parameters
$\lambda_1,\lambda_2,\sigma_1,\sigma_2,s_1,s_2$
obey three constraints
(as compared to one in the continuous case).
In particular the relation
\begin{eqnarray}
& &
\frac{1}{s_1}-\frac{1}{s_2}=- 2 a (e_\alpha-e_\gamma)^2,
\end{eqnarray}
requires $s_1$ and $s_2$ to be different
if the elliptic functions remain nondegenerate.

In order to remove the constraint $c=0$,
we have tried the assumption
(inspired by the relations (\ref{eqHalphen_system}))
\begin{eqnarray}
& &
\left\lbrace
\begin{array}{ll}
\displaystyle{
A=-a_1 \frac{\ha_\alpha(\xi+h/2)-\ha_\alpha(\xi-h/2)}{h}
      e^{\displaystyle{i(K_a \xi-\omega_a t)}},\
\xi=x-ct,\
} \\ \displaystyle{
B=-a_2 \frac{\ha_\gamma(\xi+h/2)-\ha_\gamma(\xi-h/2)}{h}
      e^{\displaystyle{i(K_b \xi-\omega_b t)}},\
      e_\gamma \not= e_\alpha,
}
\end{array}
\right.
\end{eqnarray}
but the system admits no such solution at all.

The long wave limit of the solution 
(\ref{eqcNLS_saturable_two_product}),
(\ref{eqdNLS_AL_saturable_two_sol})
is $(e_\beta-e_\alpha)(e_\beta-e_\gamma)=0$, 
and it leads to the same expression
(\ref{eqcNLS_saturable_two_product_trigo}) as in the continuous case.

\section{Discussion and conclusion}

A system of coupled discrete evolution equations,
with cubic nonlinear terms of the Ablowitz--Ladik type and
the presence of saturable nonlinearities, has been studied. 
For the single component case, 
a new type of periodic waves is formulated
in terms of the elliptic functions introduced by Halphen,
which maintain a high degree of symmetry. 
This family of periodic patterns is different from those
established earlier in the literature.

For the two-component case, solutions in terms of single
and products of elliptic functions are found,
which both correspond roughly to sinusoidal structures. 
As expected in any complicated nonlinear system,
significant constraints on the properties of the system, 
i.e.~coefficients of dispersion, cross phase modulation, and saturation,
must exist for analytical progress to be made.

A very important issue which has not been addressed in the present
work is the stability of these wave patterns. 
Computational tests must be conducted which will be performed in the near
future.

\section*{Acknowledgements}

Partial financial support has been provided by the
Research Grants Council contracts 
 HKU 711807E  and HKU 712008E.
RC acknowledges the partial support of the 
PROCORE - France/Hong Kong joint research grant
F-HK29/05T.


\vfill \eject
\section*{Appendix. Summary of solutions}

\tabcolsep=1.5truemm
\tabcolsep=0.5truemm

\begin{table}[h] 
\caption[Various considered NLS equations.]{
Various types of NLS equations considered in the text.
Types=
(discrete or continuous),
(one- or two-component),
(saturable or not),
(AL or non-AL).
Y=yes, N=no, B=both, 
ell=elliptic, tri=trigonometric.
A star (*) denotes apparently new results.
}
\vspace{0.2truecm}
\begin{center}
\begin{tabular}{| c | c | c | l | l | l | l |}
\hline 
discr/cont & components & saturable & AL & equation & solution & ref
\\ \hline  \hline 
C & 1 & Y &   & (\ref{eqcNLS_saturable})            & quadrature & \cite{GH1991}
\\ \hline  \hline 
C & 2 & Y &   & (\ref{eqcNLS_saturable_two})        & ell   & \cite{CMN2006}
\\ \hline 
  &   &   &   &                                     & 
ell (\ref{eqcNLS_saturable_two_product}), (\ref{eqcNLS_saturable_two_sol})
   & *    $c$ arb
\\ \hline 
  &   &   &   &                                     & 
ell (\ref{eqcNLS_saturable_two_product_trigo}),
    (\ref{eqcNLS_saturable_two_sol_trigo})
   & *   $c$ arb
\\ \hline 
C & 2 & Y &   & (\ref{eqcNLS_saturable_two_variant})& none  & 
\\ \hline  \hline 
D & 1 & N & N & (\ref{eqdNLS_nonAL})                & none  & 
\\ \hline  \hline 
D & 1 & N & Y & (\ref{eqdNLS_AL})                   & many  & \cite{AblowitzLadik}
\\ \hline  \hline 
D & 1 & Y & N & (\ref{eqdNLS_nonAL_saturable})      & 3sol 
(\ref{eqdNLS_nonAL_saturable_elliptic_Halphen}), 
(\ref{eqdNLS_nonAL_saturable_elliptic_Halphen_3sol}) & 
\\ \hline 
  &   &   &   &                                     & ell 
(\ref{eqdNLS_nonAL_saturable_elliptic_Halphen}), 
(\ref{eqdNLS_nonAL_saturable_elliptic_Khare}) & \cite{KRSS}
\\ \hline 
  &   &   &   &                                     & tri (\ref{eqdNLS_nonAL_saturable_dark}) & 
\\ \hline 
  &   &   &   &                                     & ell 
(\ref{eqdNLS_nonAL_saturable_elliptic_Halphen}), 
(\ref{eqdNLS_nonAL_saturable_elliptic_Halphen_sole})& *  
\\ \hline 
D & 1 & B & B & (\ref{eqdNLS_mixedAL})              & $\sech$, $\tanh$                                 & \cite{FZK1999}
\\ \hline  \hline 
D & 2 & Y & Y & (\ref{eqdNLS_AL_saturable_two})     & ell 
(\ref{eqcNLS_saturable_two_product}),
(\ref{eqdNLS_AL_saturable_two_sol})
 & *   $c=0$
\\ \hline   \hline 
\end{tabular}
\end{center}
\label{TableEqSol}
\end{table}

\label{lastpage}
\vfill\eject
\end{document}